\documentclass[aps,preprint,nofootinbib,showpacs]{revtex4}
\usepackage{epsf,graphics,graphicx}
\usepackage{amsmath}
\usepackage{amssymb,latexsym,mathrsfs}

\def\bea{\begin{eqnarray}}
\def\eea{\end{eqnarray}}
\def\ba{\begin{array}}
\def\ea{\end{array}}

\def\beq{\begin{equation}}
\def\eeq{\end{equation}}

\begin{document}

\DeclareRobustCommand{\baselinestretch{2.2}}
%\doublespacing

\title{The geometric phase and the dynamics of quantum phase transition induced by a linear quench }

%\maketitle

\author{B. Basu}
\email{banasri@isical.ac.in} %, Fax:+91(033)2577-3026}
\affiliation{Physics and Applied Mathematics Unit, Indian
Statistical Institute, Kolkata 700108, India}
\author{P. Bandyopadhyay}
\email{b_pratul@yahoo.co.in}
\affiliation{Physics and Applied Mathematics Unit, Indian
Statistical Institute, Kolkata 700108, India}

\begin{abstract}

%\begin{center}
%{\bf Abstract}
%\end{center}

%\DeclareRobustCommand{\baselinestretch{2.2}}
%\doublespacing{

 We have analysed here the role of the geometric phase in dynamical
mechanism of quantum phase transition in the transverse Ising model. We have investigated the system when it is driven at a
fixed rate characterized by a quench time $\tau_q$ across the
critical point from a paramagnetic to ferromagnetic phase.
Our argument is based on the fact that
the spin fluctuation occurring during the critical slowing down
causes  random fluctuation in the ground state geometric phase at
the critical regime.
The correlation function of the random geometric phase determines the excitation probability of the quasiparticles, which are excited during the transition from the inital paramagnetic to the ferromagnetic phase.
This helps us to evaluate the number density of the kinks formed during
the transition, which is found to scale as $\tau_q^{-\frac{1}{2}}$.
In addition, we have also estimated the spin-spin
correlation at criticality.

\end{abstract}

\maketitle

\section{Introduction}
The celebrated geometric phase, commonly known as the Berry phase
\cite{berry} arises when a quantum system is adiabatically
transported around a closed circuit in a parameter space. In this
transportation, besides the familiar dynamical phase, a geometric
phase appears when due to nonintegrability some variables fail to
return to their original values. The geometrical meaning of this
phase is precisely the holonomy in a Hermitian line bundle and
naturally defines a $U(1)$ connection \cite{simon}. The Berry
connection and the geometric phase can be defined even if the
condition of adiabaticity is relaxed \cite{aa}. Subsequently, the
generalization of the Berry phase was introduced to the non-Abelian
case \cite{zee} where a collection of degenerate states were cycled.
An analogous phase in classical mechanics was also pointed out
\cite{hannay,berry3}. The case with non-Hermitian evolution was also
studied \cite{garrison}. The interpretation \cite{kuratsuji} of the
driving parameters as dynamical variables, whose evolution was
influenced by the same geometrical objects as the Berry phase, has
also been noted. This made clear \cite{berry1} that the reaction of
the geometric phase on the parameters takes the form of an abstract
magnetic field which was explored earlier \cite{berry2}. A classic
connection was made \cite{berry4} between the Berry phase and the
geometric phase in beams of light \cite{rama}, which was formulated
for two state systems in the optical picture by Pancharatnam
\cite{pancha}. The concept of the Berry phase has now become very
important in quantum mechanics with wide applications in  various
fields \cite{shapere,book2}. In particular, geometric phases have
been associated with a variety of condensed matter and solid state
phenomena \cite{thou,resta,hatsugai,naka,bp,bps,bb}. Besides,
theoretical analyses it has been a subject of experimental
investigations  also \cite{shapere,p1,p2,p3,n1,n2,a1,zhang}.

A characteristic feature of all non-trivial geometrical  evolutions
is that it demonstrates the presence of non-analytic points in the
energy spectrum where the state is degenerate.
 It may be mentioned here that before the discovery of the geometrical phase, the sign change associated with the
 molecular electronic degeneracies was understood \cite{old1,old2}. The presence of degeneracy at some point is
 accompanied by the curvature in its immediate neighbourhood and a state evolving around a closed path can detect it.
These aspects  indicate  an intuitive connection between the
criticality around the degeneracy point and the geometric phase.
Recently, the geometric phase has been studied \cite{car,zhu,hama}
in the context of quantum phase transitions (QPT)\cite{ss,vojta} to
investigate the regions of criticality. QPTs exhibit the crossover
of ground states at a critical value of some external parameter at
zero temperature and are caused by the presence of degeneracy in the
energy spectrum. The geometric phase, which measures the curvature
associated with the degeneracy point, can contemplate the energy
structure of the system and essentially captures certain
characteristic features of QPT. Pachos and Carollo \cite{car1} have
exhaustively studied the role of geometric phase as a tool to probe
quantum phase transitions in many-body systems and suggested the
feasibility of the experimental realization also.
 The geometric phase has also been used as a topological test to reveal quantum phase transition \cite{hama}.
 Recently it has been shown that the geometric phase associated
with the ground state of the XY model exhibits universality in the critical properties and obeys the scaling behavior
near the critical point \cite{zhu}, which established a significant connection between the geometric phase and the
quantum phase transition.

The inter-relation  between the geometric phase and quantum phase
transition has so far been concerned with the detection of
criticality through this phase and focus on equilibrium scalings of
various physical quantities near the critical point. However, till
now, the role of the geometric phase in the dynamical mechanism of
QPT has not been explored.  It is our novel attempt here to show
that the geometric phase plays a crucial role in the dynamical
mechanism of the QPT when it is induced by a linear quench,  where
non-equilibrium dynamics while passing through a critical point
becomes relevant. The Kibble-Zurek  mechanism (KZM) \cite{kib,zur}
which is a universal theory of the dynamics of the second order
phase transition, has recently been explored in the zero temperature
limit to study the dynamics of the quantum phase transition . The
behaviour of the quantum Ising model in one dimension has been
investigated during a quench caused by gradual tuning of the
transverse bias field \cite{zurek05,dziar05}. The system is  driven
at a fixed rate characterized by the quench time $\tau_q$ across the
critical point from  paramagnetic to ferromagnetic phase. In
agreement with KZM (which recognizes that evolution is approximately
adiabatic far away, but becomes approximately impulse sufficiently
near the critical point), quantum state of the system after the
transition exhibits a characteristic correlation length
${\hat{\xi}}$  proportional to the square root of the quench time
$\tau_q$: ${\hat{\xi}}=\sqrt{\tau_q}$. The inverse of this
correlation length is known to determine the average density of
defects formed during  the transition. We focus here on the role of
the geometric phase in the dynamical mechanism of QPT induced by a
linear quench, and have established a nontrivial connection between
KZ mechanism and the geometric phase. It is pointed out that the
quench induced QPT in a spin chain generates spin fluctuation which
causes the randomization of the time dependent magnetic field so
that the geometric phase acquires random values. The two point
correlation of randomized geometric phase corresponds to the
excitation probability of the quasiparticles. From this we have
estimated the number density of defects generated in the transverse
Ising model during the phase transition.

The plan of the paper is as follows. In
section 2, we review the geometric phase (Berry phase) of the ground
state of the transverse Ising model as discussed in \cite{car}.
Section 3 deals with the  discussion of  the geometric phase during
critical slowing down induced by a linear quench. In section 4 we
estimate the number density of kinks from the two-point correlation
of the Berry phase factor. Section 5 is devoted to discuss the
spin-spin correlation at the critical point.

\section{Geometric phase of the ground state of the transverse Ising model}
The Hamiltonian of a one-dimensional transverse Ising model is given by
 \begin{equation}\label{h11}
H=-J(\sum_{-M}^{+M} \sigma_{i+1}^z \sigma_i^z~+\lambda\sigma_i^x)
 \end{equation}
 with $M=\frac{N-1}{2}$, where $N$ (odd) denotes the number of sites, $\lambda$ is the external
 field and $\sigma$'s are the standard Pauli matrices. We assume
 periodic boundary condition.
 In the transverse Ising model, the geometric phase of the ground state is evaluated by applying a rotation of an angle $\alpha$ around the $z$-axis to each spin \cite{car, car1}. A new set of Hamiltonians $H_\alpha$ is constructed from the Hamiltonian (\ref{h11}) as
 \begin{equation}\label{h2}
 H_\alpha=U(\alpha)~H~U^\dagger(\alpha)
 \end{equation}
 where
 \begin{equation}\label{g}
 U(\alpha)=\prod_{j=-M}^{+M} \exp(i\alpha\sigma_j^z/2).
 \end{equation}
 We use the standard Jordan-Wigner transformation, which maps the spins to one dimensional spinless fermions,
 \begin{equation}
 a_j=\left( \prod_{i<j}\sigma_i^z \right) \sigma_j^\dagger
 \end{equation}
 and the Fourier transforms of the fermionic operator
 \begin{equation}
 d_k=\frac{1}{\sqrt{N}}\sum_j ~a_j\exp\left( \frac{-2\pi j~k}{N}\right)~~~{\rm{with}}~~~ k=-M,...+M
 \end{equation}
 such that the Hamiltonian $H_\alpha$ can be diagonalised by transforming the fermionic operators in momentum space and then using the Bogoliubov transformation. This yields
 \begin{equation}\label{h3}
 H=\sum_k~2J\Lambda_k(c^\dagger_k c_k -\frac{1}{2})=\sum_{k} \epsilon_k(c^\dagger_k c_k -\frac{1}{2})
 \end{equation}
 where
 \begin{equation}
 c_k=d_k\cos\frac{\theta_k}{2}-id^\dagger_{-k}e^{2i\alpha}\sin\frac{\theta_k}{2},
 \end{equation}
 is the fermionic annihilation operator in the momentum space,
 with
 \begin{equation}
 \cos\theta_k=\frac{\cos\frac{2\pi k}{N}-\lambda}{\Lambda_k},
 \end{equation}
 and
 \begin{equation}
\displaystyle{ \Lambda_k=\sqrt{\left(\lambda-\cos\left(\frac{2\pi
k}{N}\right)\right)^2+\sin^2\left(\frac{2\pi k}{N}\right)}},
 \end{equation}
 the energy spectrum.

 The ground state  $|g>$ of the system is expressed as
 \begin{equation}\label{g}
 |g>=\prod_{k=1}^M(\cos\frac{\theta_k}{2}|0>_k|0>_{-k} -i\exp(2i\alpha)\sin\frac{\theta_k}{2}|1>_k|1>_{-k}
 \end{equation}
 where $|0>_k$ and $|1>_k$ are the vacuum and single fermionic excitation of the $k$-th momentum mode respectively.
 The geometric phase of the ground state, accumulated by varying the angle $\alpha$ from $0$ to $\pi$, is found to be \cite{car,car1}
 \begin{equation}\label{phase1}
 \Gamma=-i\int_0^\pi \left< g\left|\frac{\partial}{\partial\alpha}\right|g\right> d\alpha =\sum_{k>0}\pi(1-\cos\theta_k)
 \end{equation}

 \section{Ground-state geometric phase in  a varying magnetic field %of the transverse Ising model
 and criticality}
  We start with the review of the geometric phase of a spin $1/2$ system  in a $varying$ magnetic field. The
  Hamiltonian of a single spin $1/2$ system in the presence of an external time dependent magnetic field is given by
 \begin{equation}\label{h1}
 H(t)=\frac{1}{2}{\bf B}(t). {\vec{\sigma}}
 \end{equation}
 where ${\vec{\sigma}}=(\sigma_x,\sigma_y, \sigma_z)$ are the Pauli matrices and ${\bf{B}}(t)=B_0{\bf n}(\theta(t))$,  ${\bf n}(\theta,t)$ being the unit vector with
   ${\bf n}=(\sin\theta \cos\phi,\sin\theta\sin\phi,\cos\theta)$. If the external magnetic field is varied
   adiabatically, the instantaneous spin states %follow the directions of ${\bf n}$ and
  can be expressed in the $\sigma_z$ basis as
  \begin{equation}\label{art}
\begin{array}{ccc}
    \displaystyle{|\uparrow_n;t>}&=&\displaystyle{\cos \frac{\theta}{2} |\uparrow_z> +~ e^{i\phi} \sin
    \frac{\theta}{2}|\downarrow_z> }\\
    &&\\
    \displaystyle{|\downarrow_n;t>}&=&\displaystyle{\sin \frac{\theta}{2} |\uparrow_z> -~ e^{i\phi}\cos
    \frac{\theta}{2}|\downarrow_z>}
\end{array}
\end{equation}
where $|\uparrow_z>,|\downarrow_z>$ are the eigenstates of the
$\sigma_z$ operator.

For a cyclic time evolution i.e. for ${\bf{B}}(T)={\bf{B}}(0)$, apart from the dynamical phase, the  eigenstates
acquire a geometric phase also and we can write,
\begin{equation}
|\uparrow_{n(T)}>=e^{i\delta} e^{i\gamma_B}|\uparrow_{n(0)}>
\end{equation}
where the dynamical phase $\delta=\int_{0}^T B_0(t) dt$ and the geometric phase (GP)
$\gamma=\oint{\bf A}^\uparrow \cdot d{\vec{\lambda}}$;\\  ${\vec{\lambda}}$ is the set of control parameters and
 ${\bf A}^\uparrow=i< \uparrow_n|\nabla_\lambda|\uparrow_n>$ is the so called Berry connection. In our case,
 ${\vec{\lambda}}=(\theta,\phi)$ and a straight forward calculation shows that
 \begin{equation}
 \gamma_{\uparrow}=-\gamma_{\downarrow}=\pi(1- \cos\theta)
 \end{equation}
 It may be pointed out that though the eigenenergies depend on $B_0(t)$, the eigenstates depend on ${\bf{n}}(\theta(t))$ only.
 The GP is proportional to the solid angle subtended by ${\bf{B}}$ with respect to the degeneracy ${\bf{B}}=0$.
It may also be noted that fluctuation in the external magnetic
field will obviously induce fluctuation in the geometric phase
($\gamma_{\uparrow}$ or $\gamma_{\downarrow})$ of the corresponding
spin eigenstate  through fluctuation in $\cos\theta$ \cite{palma}.

This formulation may be extended for a chain of spin $\frac{1}{2}$
system with XY type of interactions, which exhibits QPT for a
critical magnetic field. Subsequently, we can analyse the dynamics
of the geometric phase when there is a gradual slowing down of the
external magnetic field. We consider that the quantum fluctuation
responsible for a QPT induces spin fluctuation \cite{9}  which
eventually generates random fluctuation in the ground state Berry
phase. In the critical regime, the Berry phase of the ground state
of the relevant system undergoes a random fluctuation which is
assumed to be small for a slow transition and large enough for a
fast transition. In the present communication, we have used this
specific feature of the geometric phase of the ground state of the
transverse Ising model during its critical slowing down and showed
that the  two-point correlation of the Berry phase factors may be
used to determine the excitation probability of the quasiparticles
and hence the density of defects formed during the transition.

 Let the system governed by the Hamiltonian (1)  be initially in its ground state  at the large values of $\lambda>>1$,
 a paramagnet with all spins polarized along the x-axis. Gradually, when $\lambda$ is ramped down to  very small
 values $\lambda<<1$, %through the critical point $\lambda_c=1$,
there may be two degenerate
 ground states representing ferromagnets either with all up spins or all down spins along the z-axis. For a large number of spins i.e. $N \rightarrow \infty$, the energy gap at the critical point $\lambda=\lambda_c=1$ tends to zero implying
 that the system is excited while passing (with a finite speed) through the critical point and the system settles down in the final state with $\lambda=0$.  %The spin of the quasiparticle undergoing excitation changes its orientation with the quench time $t=\tau_q$.
 As a result the system in the final state represents a quantum superposition of states as
 $$
 \displaystyle |....\downarrow \downarrow \uparrow \uparrow\uparrow\downarrow\downarrow\downarrow\downarrow\uparrow......>
 $$
with finite domains of spins pointing up or down. The domains are
separated by kinks or
 antikinks, where the polarization of spin changes its orientation. During the excitation, spin flips occur at different domains of the chain so that in these regions there are two nearest neighbour spins with opposite orientations. The average density of kinks depends on the
 transition rate. For a slow transition the number of kinks present is small.  For a fast transition,
 the orientation of individual spins  become random, uncorrelated with their nearest neighbours and the density
 of kinks becomes large. We would like to explore the situation when the time dependent field $\lambda(t)$ driving the transition is given by
 \begin{equation}\label{quench}
 \lambda(t)(t<0)=-\frac{t}{\tau_q}
 \end{equation}
 where the  quench time $\tau_q$ is an adjustable parameter.
We may note that before the quench (-$t>\tau_q)$, the system is in
the symmetric (paramagnetic) phase. During the quench time $\tau_q$,
it undergoes a transition towards the symmetry broken ferromagnetic
phase and finally at the end of the quench ($t=0$) it settles down.

The total time dependent Hamiltonian of the system may be written as
\begin{equation}
H(t)=H_{0}(t)+H_{flip}(t)
\end{equation}
If ${\cal{N}}$ pairs of quasiparticles are excited with momentum $k_{0_i}, -k_{0_i}$ with ($i=1, 2,...{\cal{N}}$) at a certain quench time $t=\tau_q$,  the time dependent spin flip interaction is given by
\begin{equation}
H_{flip}(t)=-2J \sum_{k_0=1}^{\cal{N}}\frac{\Delta_{k_0}(t)}{2}\sigma^{x_i}
\end{equation}
with $$\sigma^{x_i}=|\downarrow><\uparrow|~~~~
\rm{or}~~~|\uparrow><\downarrow|$$ and $\Delta_{k_0}(t)$ being a
time dependent dimensionless parameter characterising the spin flip
interaction strength. Here the summation includes the spins which
get excited during the transition and finally are flipped with
opposite orientations. This term corresponds to the standard
Landau-Zener Hamiltonian \cite{zener}. The first term in eqn. (17)
is given by
\begin{equation}
H_{0}(t)=-J(\sum_{n} \sum_{i=1}^{N_n}\sigma_{i}^z \sigma_{i+1}^z+\lambda(t)\sigma_i^x)
\end{equation}
$N_n$ being the number of spins between the flipped spins in the
$n$th domain and $n$ is the number of domains in the whole system.
This corresponds to the Hamiltonian of the surrounding Ising spin
chain. We
 have neglected the coupling between the flipped spins with the surrounding Ising chain.

In the process of critical slowing down certain spins which are
excited, undergo fluctuations around the z- axis.
 In fact, the spin fluctuation is instrumental
in a metal-insulator transition which in zero temperature limit
corresponds to QPT \cite{9}. When a spin gets excited, it
 passes through the process of the change of its alignment  around
the $z$-axis until at the end of the quench at t=0 it settles with
 opposite orientation. This process of the change of alignment of spins at the quench time can be depicted as a
 fluctuation of the spin when magnetic moments  appear with random orientations.

It may be noted from eqn. (15) that during critical slowing down the
fluctuation of the angle $\theta$ associated with spin fluctuation,
makes the Berry phase a fluctuating one. Indeed, when a
quasiparticle associated with a spin rotates around a closed path,
the Berry phase, which is given by the holonomy corresponds to the
number of magnetic flux lines enclosed by the loop. As during the
quench, the time dependent magnetic field is randomized due to the
spin fluctuation, the Berry phases acquired by the spin eigenstates
take random values.

 To evaluate the Berry phase acquired by the spin eigenstate associated with the quasiparticle undergoing excitation near
 the critical point we note that at the quench time $\tau_q$, spins tend to align along the z-axis as the initial paramagnetic system ($\lambda >>1$) transits to a ferromagnetic one ($\lambda <<1$) passing through the critical point ($\lambda_c=1$).
 In fact the critical Ising model corresponds to a free fermionic field theory \cite{vidal} with spins aligned along the
 z-axis.
 To determine the Berry phase at the quench time $\gamma_{k_0}(\tau_q)$ for a spin eigenstate corresponding to a
 quasiparticle with momentum $k_0$, we note that initially at $\lambda>>1$ the system is in the paramagnetic state with
 all the spins aligned along the x-axis. We consider that the time dependent magnetic field rotates around the x-axis
 with  an angular velocity $\omega_0$. At time $t=\tau_q$ ($\tau_q=\frac{2\pi}{\omega_0}$), the spin state with momentum
 mode $k_0$, picks up a geometric phase
 \begin{equation}\label{gpk}
\gamma_{k_0}(\tau_q)=\pi(1-\cos\theta_{k_0}(\tau_q))
 \end{equation}
with $\theta_{k_0}(\tau_q)=\pi/2$
 indicating that
 at  $t=\tau_q$, the geometric phase $\gamma_{k_0}(\tau_q)$ acquired by the spin eigenstate is $\pi$.

 We consider that during critical slowing down ${\cal{N}}$ pairs of quasiparticles with momentum modes
 $k_{0_i}, -k_{0_i}$ with
($i=1, 2,...{\cal{N}}$) are excited with probability $p_{k_0}$. The
corresponding spins are flipped with opposite orientations and
settle down at $t=0$ implying $\theta_{k_0}(t=0)=\pi$, such that the
random variable $\gamma_{k_0}(t=0)$ takes the value $2\pi$ with
probability $p_{k_0}$. In essence, the Berry phase acquired by a
spin eigenstate at $t=\tau_q$ and at $t=0$ are given by the random
values
\begin{equation}
 \gamma_{k_0}(\tau_q)=\pi, ~~~~ \gamma_{k_0}(t=0)=2\pi
 \end{equation}
 with probability $p_{k_0}$ and 0 with with probability $1-p_{k_0}$.
We can now estimate the number density of kinks generated in the system due to a quench induced QPT.

\section{Estimation of the number density of defects}

For our further analysis, we define a (dimensionless)  random
variable
\begin{equation}
\phi_{k_0}(t)=\frac{\gamma_{k_0}(t)}{2\pi}
\end{equation}
such that at $t=\tau_q$ and $t=0$, the Berry phase factors
$\phi_{k_0}(t)$ attain the values
\begin{equation}
 \phi_{k_0}(t=\tau_q)=1/2, ~~~~ \phi_{k_0}(t=0)=1
 \end{equation}
Let us assume  that the Berry phase factor $\phi_{k_0}(t)$  follows
the simplest stochastic differential equation \cite{21}
 \begin{equation}\label{dphi}
d{\phi_{k_0}(t)}=-\omega_{k_0}{\phi_{k_0}(t)}dt+d\eta(t)
\end{equation}
where $\omega_{k_0}$ is the frequency associated with the energy
$\epsilon_{k_0}$ of the quasiparticle near the critical point and
$\eta(t)$ denotes  the fluctuation. Let us consider $\eta(t)$ as the
Gaussian white noise with moments
\begin{eqnarray}
\displaystyle <d\eta(t)>&=& 0\\
 \displaystyle
 <d\eta(t)d\eta(t^\prime)>&=&\delta(t-t^\prime)dt^\prime
 \end{eqnarray}
Using eqn.(\ref{dphi}), and performing the averages over the
noise, the correlations of the time dependent Berry phase factors
can be evaluated \cite{nelson}. This gives
\begin{eqnarray}\label{g1}
\displaystyle <\phi_{k_0}(t)>&=& 0\\
 \displaystyle <\phi_{k_0}(t)\phi_{k_0}(t^\prime)>&=&\frac{1}{2}e^{-\omega_{k_0}(t-t^\prime)}
 \end{eqnarray}
 This leads us to derive  the two-point correlation function of the Berry phase factor $\phi_{k_0}$ during the critical
 slowing down  from $t=\tau_q$ to $t=0$ as
 \begin{equation}\label{gp1}
 \displaystyle <\phi_{k_0}(\tau_q)\phi_{k_0}(0)>=\frac{1}{2}e^{-\omega_{k_0}\tau_q}
 \end{equation}
 The correlation is large for fast transition (small $\tau_q$) and small for slow transition (large $\tau_q$).

As ${\cal{N}}$ pairs of quasiparticles are excited during the
critical slowing down with momentum modes $k_{0_i}, -k_{0_i}$
($i=1, 2,...{\cal{N}}$) with probability $p_{k_0}$, we note from relation (23) that the random variable
 $|2 \phi_{k_0}(t=\tau_q)\phi_{k_0}(t=0)|$ picks up the value 1 with
probability $p_{k_0}$ and 0
 %\end{equation}
  with probability $1-p_{k_0}$.  This suggests the excitation probability to be given by
$$p_{k_0}=2 \left< \phi_{k_0}(\tau_q)\phi_{k_0}(0)\right>$$
Using eqn.(29), this can be explicitly written as
\begin{equation}
p_{k_0}=2 \left< \phi_{k_0}(\tau_q)\phi_{k_0}(0)\right>
=e^{-\omega_{k_0}\tau_q} = e^{-2\pi
\epsilon_{k_0}\tau_q}~~~~~(\hbar=1)
\end{equation}

One should note that in the final state at t=0, the flipped spins correspond to the values of $\theta_{{k_0}}=\pi$, with $\cos\theta_{{k_0}}=-1$.   %which correspondingly give the energy expression s the critical point, For a flipped spin in Ising chain,
Now if we choose the lattice spacing $a=\frac{2\pi}{N}=1$, in the
critical region, with $\lambda=1$, the constraint
$\cos\theta_{{k_0}}=-1$ suggests from eqn. (8) that
\begin{equation}
\Lambda{_{k_0}}=1-\cos k_0
\end{equation}
The expression for the energy then yields
\begin{equation}
\epsilon_{k_0}=2J\Lambda{_{k_0}}= 2J(1-\cos k_0)= 4J\sin^2 \frac{k_0}{2} \sim  J k_0^2 ~\rm{for}~~ \rm{small}~~ k_0
\end{equation}
From this, we find
 \begin{equation}
p_{k_0}=e^{-2\pi~ J~ k_0^2\tau_q}~~~~(\hbar=1)
\end{equation}
 The number of kinks can now be
evaluated as
\begin{equation}
2{\cal{N}}=\sum_{k_0}p_{k_0} \end{equation}
For large number of spins, i.e. in the thermodynamic limit as $N\rightarrow \infty$, the expectation value of the density of kinks is given by
\begin{equation}
n=_{N\rightarrow
\infty}\frac{2\cal{N}}{N}=\frac{1}{2\pi}\int_{-\pi}^{\pi}p_{k_0}dk_0
=\frac{1}{2\pi}(2~J~\tau_q)^{-\frac{1}{2}}
\end{equation}
which shows that the density of kinks scales as
$\tau_q^{-\frac{1}{2}}$. In this novel framework, the dynamics of
the geometric phase associated with the flipped spins helped us to
estimate the density of defects produced in a system due to QPT and
the result is found to be in concordance with that derived by a
standard technique \cite{dziar05}.

\section{Spin -spin correlation at the critical point}
 We can now derive the spin-spin correlation at the critical point in terms of the space dependent correlation of the geometric phase. In the initial ground state, with $\lambda\rightarrow \infty$, due to symmetry $<\sigma^z>=0$ and the transverse magnetization $<\sigma^x>=1$. For a slow quench,  the final magnetization, $<\sigma^x> \rightarrow 0$, and in the final state at
$\lambda=0$, $<\sigma^x> = 0$. It may be noted that for a certain quench time $\tau_q$, $<\sigma^x>$ will depend on the density of kinks . In the region between  a kink-antikink pair, the direction of the spin orientation remains the same corresponding to the direction of orientation of the flipped spin. If the kink-antikink pair maintains more or less same distance from each other, the system represents a kink-antikink chain with a lattice constant $\sim$ ${\hat{\xi}}$, where ${\hat{\xi}}=\sqrt{\tau_q}$ is the Kibble-Zurek correlation length. Indeed the expected value
$<\sigma_i^x>$ for a particular quench time $\tau_q$ is \cite{cincio}
\begin{equation}
 <\sigma_i^x> \sim \frac{1}{2\pi~\sqrt{2\tau_q}}~~~(J=1)
 \end{equation}
 which is valid for $\tau_q>>1$.
 We have just analysed how the variation in the direction of the spin orientation during  the quench time
 $\tau_q$ is reflected in the geometric phase factor $\phi_{k_0}(\tau_q)$. In fact, the direction of the spin
 orientation at the spatial position  $r$ denoted by site $i$ near the critical point determines the
 geometric phase $\gamma(r)$ acquired by the spin eigenstate  after a cyclic evolution.  If $r^\prime$ denotes the
 spatial coordinate of the spin at site $i+R$, then we can define the two-point correlation of the Berry phase factor
 near the critical point as
 \begin{equation}\label{cor}
 C^{xx}_R=
 <\phi(r)~\phi(r^\prime)>
 \end{equation}
 where $\phi(r)=\frac{1}{2\pi}\gamma(r)$.
 The correlation (\ref{cor}) effectively corresponds to the the  spin-spin correlation at the critical point.
 As already mentioned,  the Berry phase undergoes  a stochastic fluctuation near the critical point. In terms of spatial variable, we consider that the Berry phase factor corresponding to the spin state at position $r$ follows the stochastic differential equation
\begin{equation}
d{\phi(r)}=-\frac{\phi(r)}{{\hat{\xi}}}dr+d\eta(r)
\end{equation}
Here $\eta(r)$ is the fluctuation and ${\hat{\xi}}$ is the KZ correlation length which is the relevant length scale in the system. We consider that $\eta(r)$ represents Gaussian distribution with white noise and satisfies the moments
\begin{eqnarray}
\displaystyle <d\eta(r)>&=& 0\\
 \displaystyle <d\eta(r)d\eta(r^\prime)>&=&\delta(r-r^\prime)dr^\prime
 \end{eqnarray}
From this the correlation function of the Berry phase factor can be
evaluated \cite{nelson} and is given by
\begin{eqnarray}
\displaystyle <\phi(r)>&=& 0\\
 \displaystyle <\phi(r)\phi(r^\prime)>&=&\frac{1}{2}e^{-\frac{(r-r^\prime)}{{\hat{\xi}}}}
 \end{eqnarray}
 This implies that
 \begin{equation}
 C_R^{xx}=<\phi(r)\phi(r^\prime)>=\frac{1}{2}e^{-
 \frac{R}{{\hat{\xi}}}}=0.5 e^{-
 \frac{R}{{\hat{\xi}}}} ~~~~~(\hbar=J=1)
 \end{equation}
  This  can be compared with the numerical estimate of spin -spin correlation at the critical point in ref.[48]
\begin{equation}
 C_R^{xx}\approx \frac{0.44}{\tau_q}e^{-2.03
 \frac{R}{{\hat{\xi}}}}
 \end{equation}
which is valid for $\tau_q >>1$ and $R>>{\hat{\xi}}$

It is noted that in contrast with the result (43), the relation (42)
is a generalised one and does not involve any restriction on the
values of $\tau_q$ and $R$. It is observed that for very small
$\tau_q$ ($\tau_q<<1)$ the spin -spin correlation tends to zero
which is consistent with the fact that for fast transition the spins
behave almost uncorrelated. Moreover, contrary to eqn. (43) we have
found that there is a finite correlation for large $\tau_q$ and
large $R$.

\section{Conclusion}
The KZ mechanism in the second order phase transition suggests that
the formation of defects is driven by thermal fluctuation. In the
analysis of spin models, the KZ  theory has been extended to study
quench induced quantum phase transition driven by quantum
fluctuation. Equivalently, we have argued that when the KZ mechanism
is encompassed in the quantum phase transition of spin systems, the
defect formation is driven by the fluctuation of the Berry phase
factor associated with the ground state. Recently, we have studied a
XY spin chain in a slowly varying time dependent magnetic field and
investigated the behaviour of the geometric phase during a linear
quench caused by a gradual turning off of the magnetic field
\cite{pla10}. In the present formulation, we have argued that in
quantum phase transition induced by a quench the Berry phase factor
undergoes a random fluctuation at criticality. The two-point
correlation function of the Berry phase factor determines the
dynamics of the QPT induced by a quench and controls the formation
of defects. It is found that the  number of kinks generated in this
transition  scales as $\tau_q^{-\frac{1}{2}}$ where $\tau_q$ is the
quench time. The result is in agreement with that obtained through
other formulations. Besides, the space dependent correlation of the
random Berry phase gives us the estimation of the spin-spin
correlation at the critical point. It is hoped that this work may
initiate investigations in relation to the role of the geometric
phase in the dynamical mechanism of quantum phase transition.

\end{document}